\documentclass[aps,pra,twocolumn,groupedaddress,amsmath,amssymb]{revtex4-1}
\usepackage{graphicx}  
\usepackage{dcolumn}   
\usepackage{bm}        
\usepackage{verbatim}   
\usepackage{hyperref} 
\usepackage[autostyle]{csquotes}

\begin{document}

\title{Entangled vacuum state for accelerated observers}
\author{Leyli Esmaelifar}
   \author{Behrouz Mirza  }
  \email{b.mirza@cc.iut.ac.ir}
   \author{Zahra Ebadi}
  \email{z.ebadi@uma.ac.ir}

\affiliation{$\dagger$Department of Physics, Institute for Advanced Studies in Basic Science (IASBS), Zanjan, Iran }
   \affiliation{ $*$Department of Physics, Isfahan University of Technology, Isfahan 84156-83111, Iran}
    \affiliation{ $\ddag$ Department of Physics, University of Mohaghegh Ardabili, P.O. Box 179, Ardabil, Iran}

\begin{abstract}
Entanglement of Dirac fields has been studied and it is known to decrease with increasing acceleration when a quantum state is shared between users in non-inertial frames. A new form of an entangled vacuum state observed by the accelerated observer is postulated in which it is assumed that entanglement is present between the modes of the
quantum field with sharp and opposite momenta defined in two causally disconnected
regions of space-time. We find that this assumption does not affect the entanglement of the system.
\end{abstract}


\maketitle


\section{\label{sec:level1}Introduction}
 Quantum entanglement and other correlations are widely used in quantum information theory \cite{Nielsen2010quantum,wilde2013quan}. The last decade has witnessed  a wide array of research efforts focused especially on quantum entanglement between non-inertial systems in quest of a more profound understanding of relativistic and quantum information. The features of the systems considered are used in the study of the information paradox of black holes \cite{eisert2010j,lloyd2006almost} and entropy production in the expanding universe \cite{ball2006entanglement,fuentes2010entanglement,friis2013entanglement,liu2016quantum,mohammadzadeh2015entanglement,mohammadzadeh2017entropy}.
For instance, one such interesting feature is that, for a uniformly accelerated observer, a communication horizon exists that does not allow him to access the whole space-time, thereby leading to both information loss and entanglement degradation \cite{alsingfu,carroll2005spacetime,mehri2011pseudo,mehri2015quantum,fuentes2005alice,alsing2006entanglement}.

The study of entanglements between two modes of a Dirac field for two relatively accelerated observers has shown that entanglement degrades with increasing relative acceleration of observers. In order to measure the entanglement, two maximally entangled modes, $k_{A}$ and $k_{R}$, of a free Dirac field are considered in a state depicted by (\ref{eq1a}) below:
\begin{equation} \label {eq1a}
\mid\phi_{k_{A},k_{R}}\rangle =\frac{1}{\sqrt{2}}\Big[\mid 0_{k_{A}}\rangle^{+}_{M}\mid 0_{k_{R}}\rangle^{+}_{M}+\mid 1_{k_{A}}\rangle^{+}_{M}\mid 1_{k_{R}}\rangle ^{+}_{M}\Big],
\end{equation}
where, the modes $k_{A}$ and $k_{R}$ are detected by Alice, the inertial observer, and Rob, the accelerated observer, respectively. The sign $+$ (or $-$) is used to indicate the particle (or anti-particle) states and $M$ is a notational shorthand for Minkowski states. Since one of the observers, Rob, is accelerated, we might expand its states in the Rindler coordinate and study the entanglements between the modes for different values of acceleration. As already mentioned above, a Rindler observer does not have access to the whole space-time; hence, a degradation is expected in the entanglement.
 In the infinite acceleration limit, entanglement between bosonic modes \cite{fuentes2005alice} tends to zero. However, an entanglement of Dirac spinor modes never vanishes altogether \cite{alsing2006entanglement}.
\\Previous study has shown that entanglement is independent of acceleration for the specific state entangled in the helicity part \cite{harsij}. We generalize the previously introduced postulation of the vacuum state \cite{fuentes2005alice,alsing2006entanglement,unruh1984rindler,atsushi2017vacuum} and introduce a new ansatz for the vacuum state observed by Rob. We assume that vacuum is an entangled sate with respect to sharp and opposite momenta.
 In order to examine other correlations of this system and see how they behave, we also compute the purity and coherency.

The rest of the paper is organized as follows. In Section \ref{sec:level2}, we briefly review previous studies of entanglement of Dirac fields and study the Unruh effect. In Section \ref{sec:level3}, a new form of entangled vacuum state is introduced and the creation and annihilation operators are obtained. Using the density matrix, we evaluate the entanglement entropy and negativity of this system. Finally, a summary of our results is provided in Section \ref{sec:level4}. Purity and relative entropy of coherence are computed in Appendices A and B, respectively.
\section{\label{sec:level2}Entanglement of Dirac fields}
\begin{figure}[h]
\centering
\includegraphics[width=0.4\textwidth]{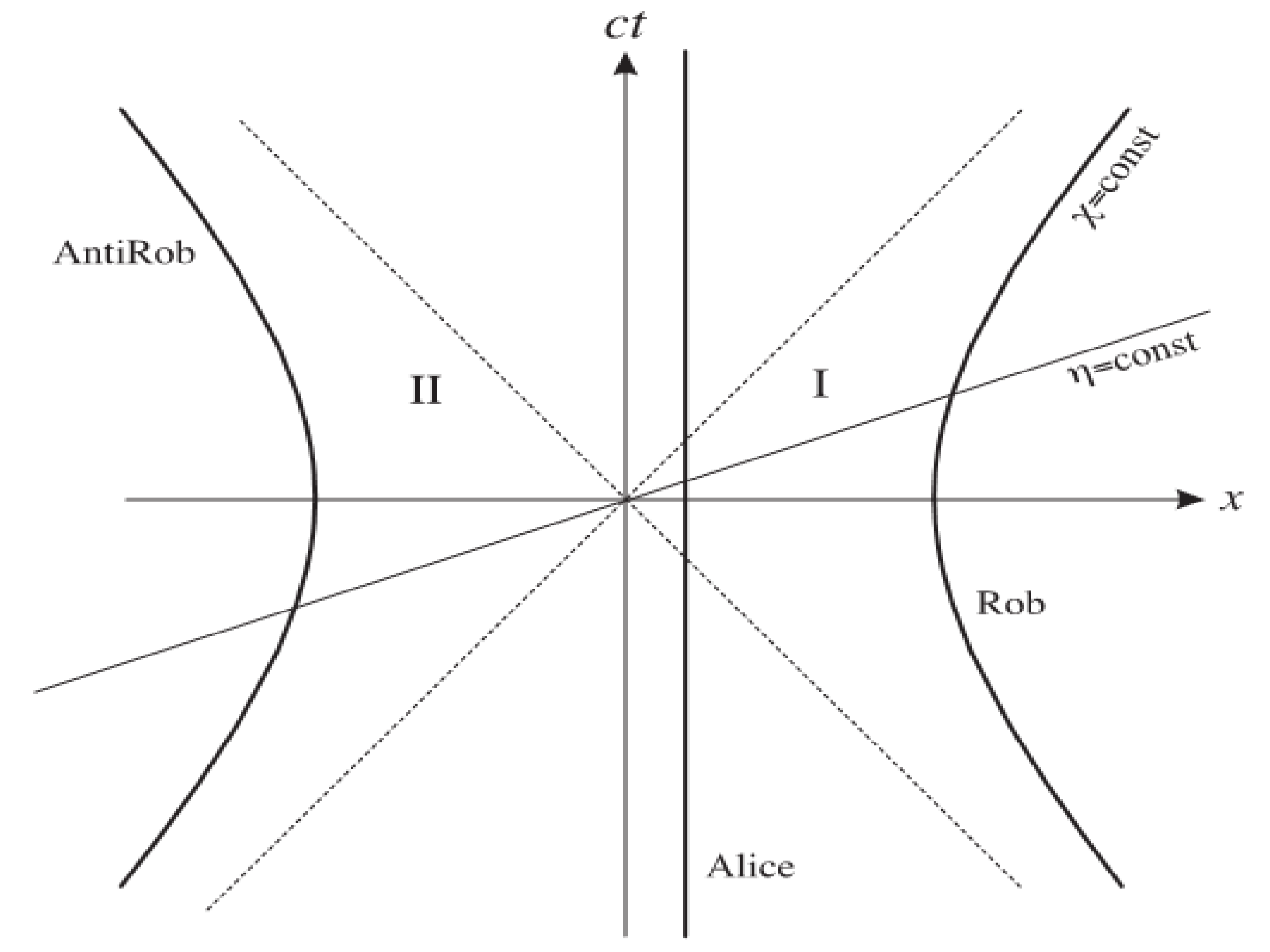}
\caption[]{\it  Rindler space-time diagram: lines of constant po-
sition $\chi = const$ are hyperbolae and all curves of $\eta=const$
are straight lines that run through the origin. Rob, the uniformly accelerated observer, travels along a hyperbola constrained
to the region $I$, and anti-Rob travels in region $II$.} \label{fig:0}
\end{figure}

Consider a free Minkowski Dirac field, $\phi$, in $1+1$ dimensional
flat space-time satisfying the Dirac equation:
\begin{equation}\label{eq1b}
i\gamma^{\mu}\partial_{\mu}\phi-m\phi=0,
\end{equation}
where, $\gamma^{\mu}$ are the Dirac-Pauli matrices and $m$ is a shorthand for the particle mass. We can expand the field in terms of Minkowski solutions of the Dirac equation:
\begin{equation}\label{eq2b}
\phi=\int dk(\hat{a}_{k}\phi^{+}_{k}+\hat{b}_{k}^{\dag}\phi^{-}_{k}),
\end{equation}
where, $\phi^{+}$ and $\phi^{-}$ are positive and negative energy solutions (fermions and anti-fermions), respectively. The operators $\hat{a}_{k}$ and $\hat{a}^{\dagger}_{k}$ are the annihilation and creation operators for the positive solutions of momentum $k$ while $\hat{b}_{k}$ and $\hat{b}^{\dagger}_{k}$ are the annihilation and creation operators for the negative energy solution of Minkowski space. These operators satisfy the usual anti-commutation relations:
\begin{equation}\label{eq3b}
\{\hat{a}_{i},\hat{a}^{\dagger}_{j}\}=\{\hat{b}_{i},\hat{b}^{\dagger}_{j}\}=\delta_{ij}.
\end{equation}
As it is the objective of the present work to study Dirac fields in non-inertial frames, use is made of the Rindler coordinate the better suits accelerated frames. As can be seen in Fig.(\ref{fig:0}),two sets of the Rindler coordinate are required for the two different regions, $I$ and $II$, in order to cover the whole Minkowski space-time.
\begin{eqnarray} \label{eq4b}
at=e^{a\chi}sinh(a\eta),\quad ax= e^{a\chi}sinh(a\eta): I,& \nonumber\\
at=-e^{a\chi}sinh(a\eta),\quad ax=-e^{a\chi}cosh(a\eta): II,
\end{eqnarray}
where, $a$ represents proper acceleration. The accelerated observer confined in region $I$ is called Rob and that in region $II$ is termed anti-Rob. Since these two regions are causally disconnected, an accelerated particle can propagate only in one of the two Rindler wedges. Therefore,
each of the two regions $I$ and $II$ has its unique quantization procedure with the corresponding solutions $\{\phi^{I+},\phi^{II+}\}$ and $\{\phi_{k}^{I-},\phi_{k}^{II-}\}$. Using these modes, the Dirac field may be expanded in the Rindler coordinate as in Eq.(\ref{eq5b}) below:
\begin{equation}\label{eq5b}
\phi=\int dk(\hat{c}^{I}_{k}\phi^{I+}_{k}+\hat{c}^{II}_{k}\phi^{II+}_{k}+\hat{d}_{k}^{I\dag}\phi^{I-}_{k}+\hat{d}_{k}^{II\dag}\phi^{II-}_{k}),
\end{equation}
where, $(\hat{c}^{\sigma\dagger}_{k},\hat{c}^{\sigma}_{k})$ represent the creation and annihilation operators for the fermions in the region $\sigma\in\{I,II\}$ of the Rindler space-time and could be determined using the Rindler orthogonality relation $\hat{c}^{\sigma}_{k}=(\psi_{k}^{\sigma +},\psi)$. The creation and annihilation operators for anti-fermion with negative modes in region $II$ of the Rindler space-time are $ (\hat{d}^{I\dagger}_{k},\hat{d}^{II}_{-k})$.
Considering Eqs.(\ref{eq2b}) and (\ref{eq5b}), one can find a relation as in (\ref{eq:3}) below between the creation and annihilation operators of both the Minkowski and Rindler space-times in the single-mode approximation using the Bogoliubov transformation \cite{takagi1986vacuum,jauregui1991dirac,mcmahon2006dirac,soffel1980dirac}:
\begin{equation}\label{eq:3}
\hat{a}^{\dagger}_{k}=\cos(r)\hat{c}^{I\dagger}_{k} - \sin(r)e^{i\varphi} \hat{d}^{II}_{-k},
\end{equation}
where, $\varphi$ is a phase that could be absorbed in the operators. $\omega$ is the frequency of the modes observed by Alice  and Rob ($\omega_{A}=\omega_{R}=\omega$) 
and $r$ is related to the acceleration $a$ by $ \tan(r)=\exp(\frac{-\pi c\omega}{a})$.
Since acceleration goes from zero to infinity, $r$ takes values in the interval $[0,\frac{\pi}{4}]$.
\\In a previous study \cite{fuentes2005alice}, the Minkowski particle vacuum in terms of the Rindler Fock space is postulated for the mode k to be of the following form:
\begin{equation}\label {eq:2}
\mid 0_{k}\rangle^{+}=\sum_{n=0}^{1} A_{n}\mid n_{k}\rangle_{I}^{+}\mid n_{-k}\rangle_{II}^{-}.
\end{equation}
We should note that Minkowski vacuum is direct product as $\vert 0\rangle=\prod_{k,k'}\vert0_k\rangle^{+}\vert0_{k'}\rangle^{-}$. For simplicity, we assume Alice's detector is only sensitive to one mode $k_A$ and Rob's detector only detects the mode $k_R$. Now we can focus on  $\mid 0_{k}\rangle^{+}$. As we can see the vacuum state in Eq.(\ref{eq:2}) has been decomposed as a tensor product of particle modes in region $I$ and antiparticle modes in region $II$.
 \\It should be noted that Eq.(\ref{eq:2}) is not entangled with respect to $k$ and $-k$. The coefficient $A_{n}$ in Eq.(\ref{eq:2}) may be obtained using $\hat{a}_{k}\mid0_{k}\rangle^{+}=0$ and the normalization of $\mid0_{k}\rangle^{+}$. Thus, the vacuum state for a certain momentum mode of Minkowski space in terms of the Rindler state will be as follows
\begin{equation} \label {eq:4}
\mid 0_{k}\rangle^{+}=\cos(r)\mid 0_{k}\rangle_{I}^{+}\mid0_{-k}\rangle_{II}^{-} + e^{-i\varphi} \sin(r)\mid 1_{k}\rangle_{I}^{+}\mid1_{-k}\rangle_{II} ^{-},
\end{equation}
Using the creation operator, the excited Minkowski state may be found in terms of the Rindler state as in Eq.(\ref{eq:5}) below:
\begin{equation} \label {eq:5}
\mid1_{k}\rangle^{+}=\hat{a}^{\dagger}_{k}\mid0_{k}\rangle^{+}=\mid1_{k}\rangle_{I}^{+}\mid0_{-k}\rangle_{II}^{-}.
\end{equation}
This yields the expansions of $\mid 0_{k}\rangle^{+}$ and $\mid 1_{k}\rangle^{+}$ in the Rindler coordinate as in Eqs.\eqref{eq:4} and \eqref{eq:5}, respectively, so instead of $\mid 0_{k_{R}}\rangle ^{+}$ and $\mid  1_{k_{R}}\rangle^{+}$ in Eq.(\ref{eq1a}), Eqs.(\ref{eq:4}) and (\ref{eq:5}) may be employed  , respectively. It is, now, straightforward to compute the density matrix and study the entanglement of the system.
\subsection{Unruh effect}
 We study the Unruh effect with this vacuum state, and
find the Unruh temperature.
 The Unruh effect is an amazing result of quantum field theory that predicts the uniform accelerated observer surprisingly witnessing a thermal bath with a temperature related to the acceleration as follows \cite{unruh1976notes,davies1975scalar,carroll2004spacetime}:
\begin{equation} \label {eq:1}
T=\frac{\hbar a}{2\pi K_{B}c}.
\end{equation}
 In fact, when the observer accelerates in region $I$, his detector finds a number of particles that may be evaluated as follows:
\begin{eqnarray}
\label {eq:7a}
n^{+}_{k}&=&^{+}\langle 0_{k}\mid \hat{c}^{k\dagger}_{I} \hat{c}^{k}_{I}\mid 0_{k}\rangle ^{+}
\nonumber\\
&=& \sin^{2}(r)=\frac{1}{e^{\hbar \omega /K_{B}T}+1},
\end{eqnarray}
where, $T$ is temperature as defined in Eq.(\ref{eq:1}). Equation (\ref{eq:7a}) is a thermal Fermi-Dirac (FD) distribution of particles detected by the  accelerated observer.
\section{\label{sec:level3}New entangled vacuum state for Dirac fields}
As mentioned earlier, the entanglement between the modes of a Dirac field observed by Alice and Rob degrades when the relative acceleration increases. This is because Rob lacks access to the whole space-time; more precisely, he is confined to region $I$ and has no access to region $II$. Now we would like to see what will happen if we consider that the vacuum state is entangled.
Here, instead of postulating the previous vacuum state (Eq.(\ref{eq:2})), we might postulate that the vacuum is an entangled state with respect to modes $k$ and $-k$ as captured by (\ref{eq7c}) below:
\begin{equation} \label{eq7c}
\mid 0_{k}\rangle^{+}=\sum_{n=0}^{1}\Big[ A_{n}\mid n_{k}\rangle_{I}^{+}\mid n_{-k}\rangle_{II}^{-}+B_{n}\mid n_{-k}\rangle_{I}^{+}\mid n_{k}\rangle_{II}^{-}\Big],
\end{equation}
This state is an entangled state with respect to modes $k$ and $-k$. Entanglement of vacuum has been investigated in other studies such as \cite{reznik}and it has been shown that there is entanglement between the modes in region $I$ and $II$, which means that we can consider a state as in Eq.(\ref{eq7c}).
\\In Eq.(\ref{eq7c}), $ A_{n}$ and $ B_{n}$ are two coefficients which will be obtained later.
 Here we have more than one single mode $k$, and we use the following form of the creation operator, which is a generalized form of the earlier creation operator in Eq.(\eqref{eq:3}).
\begin{equation}\label{eq:8}
\hat{a}_{k}^{\dagger}=\sum_{k}\cos(r)\hat{c}^{\dagger}_{k,I}-e^{i\varphi} \sin(r)\hat{d}_{k,II},
\end{equation}
This satisfies the usual anti-commutation rule. Using the relations $\hat{a}_{k}\mid0_{k}\rangle^{+}=0$, $\langle0_{k}\mid0_{k}\rangle^{+}=1$, and Eq.\ref{eq:8}, the coefficients $A_{n}$ and $B_{n}$ in Eq.(\ref{eq7c}) could be found. Therefore,
\begin{eqnarray} \label {eq:7}
&\mid& 0_{k}\rangle^{+}=\\\nonumber&
 & \alpha \Big[ \cos(r)\mid0_{k}\rangle_{I}^{+}\mid0_{-k}\rangle_{II}^{-}+ e^{-i\varphi} \sin(r)\mid 1_{k}\rangle_{I}^{+}\mid1_{-k}\rangle_{II}^{-} \Big] \\ & &+\beta \Big[ \cos(r) \mid0_{-k}\rangle_{I}^{+}\mid 0_{k}\rangle_{II}^{-}+ e^{-i\varphi} \sin(r) \mid 1_{-k}\rangle_{I}^{+}\mid 1_{k}\rangle_{II}^{-} \Big] \nonumber,
\end{eqnarray}
where, $\alpha$ and $\beta$ are two coefficients added for normalization and $\alpha^{2} + \beta^{2}=1$. When $\alpha=1$ and $\beta=0$, the previous case is obtained as in Eq.(\ref{eq:4}). It should be noted that this postulation is different from that of the beyond single mode analysis \cite{bruschi2010unruh}.\\
 We are now in a position to show that the well-known Unruh effect is valid when this new ansatz is considered for the vacuum state. In fact, when the observer in region $I$ accelerates, his detector detects a number of particles that may be evaluated as follows:
\begin{eqnarray}
\label {eq:13}
n^{+}_{k}&=&^{+}\langle 0_{k}\mid \hat{c}^{\dagger}_{k,I} \hat{c}_{k,I}+\hat{c}^{\dagger}_{-k,I} \hat{c}_{-k,I}\mid 0_{k}\rangle ^{+}
\nonumber\\
&=& ( \alpha ^{2} +\beta ^{2})\sin(r)^{2}=\frac{1}{e^{\hbar \omega /K_{B}T}+1},
\end{eqnarray}
where, $T$ is temperature as defined in Eq.(\ref{eq:1}). Again, Equation (\ref{eq:13}) is a thermal Fermi-Dirac (FD) distribution of particles detected by the accelerated observer similar to the previous cases and is known as the Unruh effect.
\\Applying $\hat{a}_{k}^{\dagger}$ to Eq.(\ref{eq:4}) again yields Eq.(\ref{eq:5}). However, we apply it to Eq.(\ref{eq:7}) to obtain the expansion of one Minkowski fermion state in terms of the Rindler states as follows:
\begin{equation} \label{eq9c}
\mid1_{k}\rangle^{+}=\alpha\mid1_{k}\rangle^{+}_{I}\mid0_{-k}\rangle_{II}^{-}+ \beta \mid1_{-k}\rangle_{I}^{+}\mid0_{k}\rangle_{II}^{-},
\end{equation}
\\which is an entangled state of a fermionic mode in two causally disconnected regions, $I$ and $II$. Now, we can study the entanglement of the system.
\subsection{Density matrix for fermions in non-inertial frames}
\begin{figure}[h]
\centering
\includegraphics[width=0.4\textwidth]{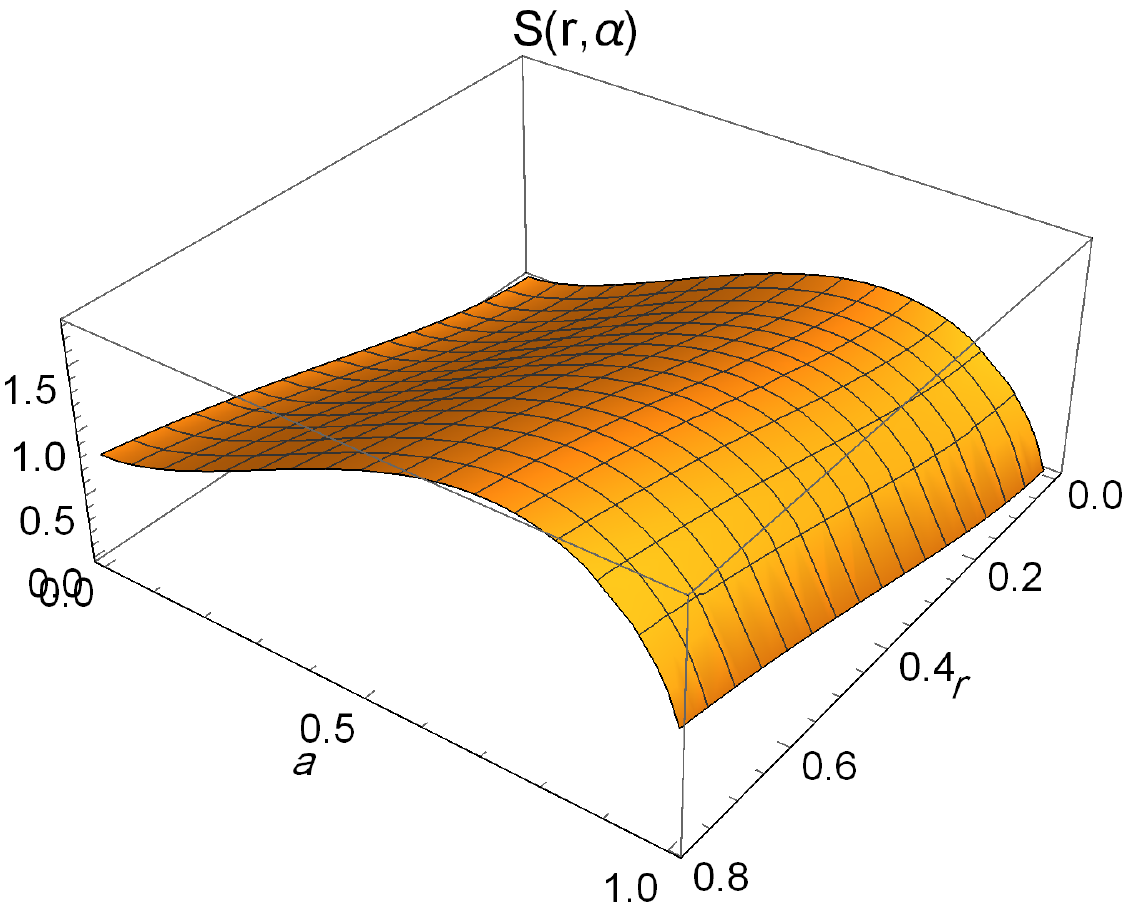}
\caption[]{\it  Entanglement entropy as a function of $\alpha$ and $r$.} \label{fig:1}
\end{figure}

In order to measure the entanglement of fermions in non-inertial frames in these protocols, we consider a Bell state in the form of Eq.(\ref{eq1a}) and replace Rob's state's, $\mid0_{k_{R}}\rangle$ and $\mid1_{k_{R}}\rangle$, by those that obtained in Eqs.(\ref{eq:7}) and (\ref{eq9c}), respectively. The density matrix of Alice, Rob, and anti-Rob can be defined as $\rho_{A,I,II}=\mid\phi_{k_{A},k_{R}}\rangle\langle\phi_{k_{A},k_{R}}\mid$. As regions $I$ and $II$ are causally disconnected, we need to trace out over the modes in region II in order to find the reduced density matrix connecting Alice and Rob, $\rho_{A,I}$:
\begin{widetext}
\begin{eqnarray}
\rho_{A,I}&&=Tr_{II}(\rho_{A,I,II})=\nonumber\\ &
&\frac{\beta^{2}\cos(r)}{2}\mid0_{A},0_{-k}\rangle\langle1_{A},1_{-k}\mid +\frac{\beta^{2} \cos(r)}{2}\mid1_{A},1_{-k}\rangle\langle 0_{A},0_{-k}\mid +\frac{\alpha^{2}\cos^{2}(r)}{2}\mid 0_{A},0_{k}\rangle\langle0_{A},0_{k}\mid
\nonumber\\ && \frac{\beta^{2}\cos^{2}(r)}{2}\mid 0_{A},0_{-k}\rangle\langle0_{A},0_{-k}\mid + \frac{\alpha^{2}\sin^{2}(r)}{2}\mid 0_{A},1_{k}\rangle\langle0_{A},1_{k}\mid + \frac{\beta^{2}\sin^{2}(r)}{2}\mid 0_{A},1_{-k}\rangle\langle0_{A},1_{-k}\mid +
\nonumber\\ &&\frac{\alpha^{2}\cos(r)}{2}\mid1_{A},1_{k}\rangle\langle0_{A},0_{k}\mid + \frac{\alpha^{2}\cos(r)}{2}\mid0_{A},0_{k}\rangle\langle1_{A},1_{k}\mid + \frac{\beta^{2}}{2}\mid1_{A},1_{-k}\rangle\langle1_{A},1_{-k}\mid +\frac{\alpha^{2}}{2}\mid1_{A},1_{k}\rangle\langle1_{A},1_{k}\mid .
\end{eqnarray}
\end{widetext}
 The reduced density matrix can be written in the basis $\mid 0_{A},0_{k}\rangle$, $\mid 0_{A},0_{- k}\rangle$,$\mid 0_{A},1_{k}\rangle $,$\mid 0_{A},1_{- k}\rangle $,$\mid 1_{A},0_{k}\rangle$,$\mid 1_{A},0_{- k}\rangle$,$\mid 1_{A},1_{k}\rangle$,$\mid 1_{A},1_{-k}\rangle$ as follows
\begin{widetext}
\begin{eqnarray}
\label {eq:11}
 \rho_{A,I}=\left(
 \begin{array}{cccccccc}
\frac{\alpha^{2}\cos^{2}(r)}{2} & 0 & 0 & 0 &0& 0&\frac{\alpha^{2}}{2}\cos(r)&0 \\
0 & \frac{\beta^{2}\cos^{2}(r)}{2} & 0 & 0&0&0&0&\frac{\alpha^{2}\cos(r)}{2} \\
0&0&\frac{\alpha^{2}\sin^{2}(r)}{2}&0&0&0&0&0\\
0&0&0&\frac{\beta^{2}\sin^{2}(r)}{2}&0&0&0&0\\
0&0&0&0&0&0&0&0\\
0&0&0&0&0&0&0&0\\
\frac{\alpha^{2}\cos(r)}{2} & 0 & 0 & 0 &0&  0&\frac{\alpha^{2}}{2}&0 \\
0 & \frac{\beta^{2}\cos(r)}{2}& 0 & 0 & 0 & 0 & 0 & \frac{\beta^{2}}{2} \end{array}
 \right )  .
\end{eqnarray}
\end{widetext}
The eigenvalues of $\rho_{A,I}$ are
\begin{eqnarray}
 \lambda(\rho_{A,I})=&\lbrace  0,0,0,0,\frac{(1-\alpha ^{2})}{2}\sin^{2}(r), \frac{\alpha^{2}}{2}\sin^{2}(r)\nonumber\\ &,\frac{\alpha ^{2}}{4}(3 + \cos(2r)),\frac{(1-\alpha ^{2})}{4}(3 + \cos(2r))\rbrace ,
 \end{eqnarray}
where, we used $\beta^{2}=1-\alpha^{2}$. Thus, the entanglement between the mode II and (modes A and I) is depicted in Fig.(\ref{fig:1}) since $S(\rho(II))=S(\rho(A,I))=-\sum_{i=1}^{8} \lambda_{i}log_{2}\lambda_{i}$.

\begin{figure}[h]
\centering
\includegraphics[width=0.4\textwidth]{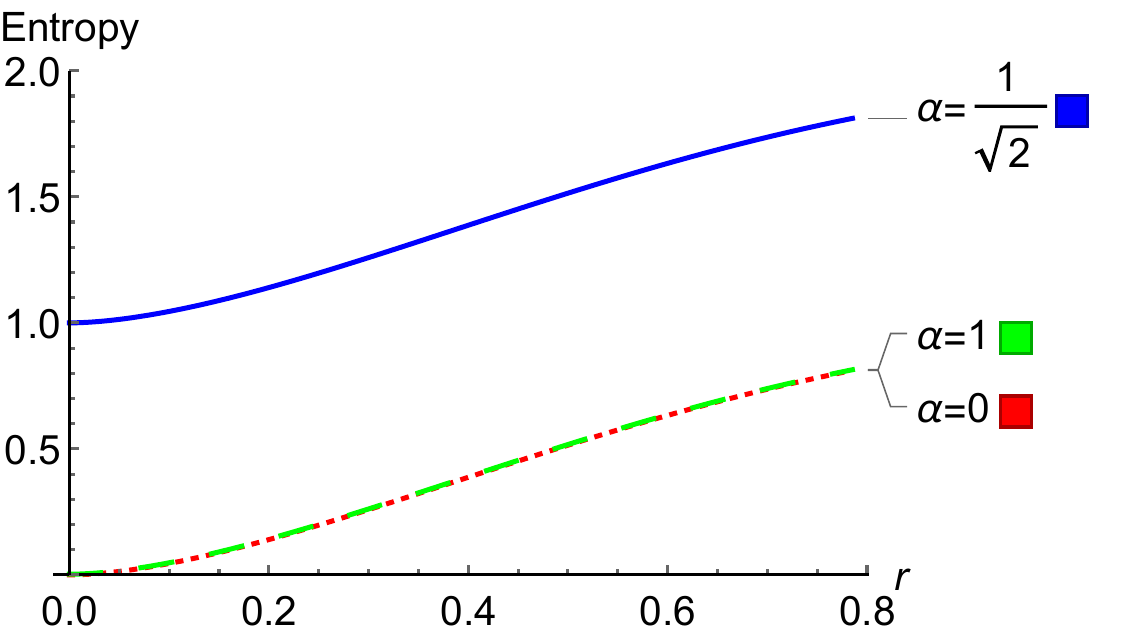}
\caption[]{\it (Color online) Bipartite pure state entanglement entropy as a funtion of $r$ for some values of $\alpha$.\\Dashed curve (Green) for $\alpha=1,\beta=0$. Thick solid curve (Blue) for $\alpha=\beta=\frac{1}{\sqrt{2}}$. Dotted curve (Red) for $\alpha=0$,  $\beta=1$.} \label{fig:2}
\end{figure}
As we can see in Figs.(\ref{fig:1}) and (\ref{fig:2}), the entanglement entropy will have its maximum value
 when $\alpha=\beta=\frac{1}{\sqrt{2}}$. It can be argued that, when $\alpha=\beta=\frac{1}{\sqrt{2}}$, both $\mid n_{k}\rangle_{I}\mid n_{-k}\rangle_{II}$ and $\mid n_{-k}\rangle_{I}\mid n_{k}\rangle_{II}$ modes exist so the information from each mode will be less than the case in which $\alpha=1$ (or $\alpha=0)$ where we only have the mode $\mid n_{k}\rangle_{I}\mid n_{-k}\rangle_{II}$ ( or $\mid n_{-k}\rangle_{I}\mid n_{k}\rangle_{II}$). Thus, as expected, the entanglement entropy for the maximally entangled case $ \alpha=\beta=\frac{1}{\sqrt{2}}$ is maximum. As we can see in Fig.(\ref{fig:2}), the entanglement entropy increases with rising acceleration. The reason is that entanglement entropy is not a suitable measure to use in mixed states \cite{entanglement2002Dagmare}. We, therefore, need to use other measures to explore entanglement of the system.
\subsection{\label{sec:level3}Negativity}
\begin{figure}[h]
\centering
\includegraphics[width=0.4\textwidth]{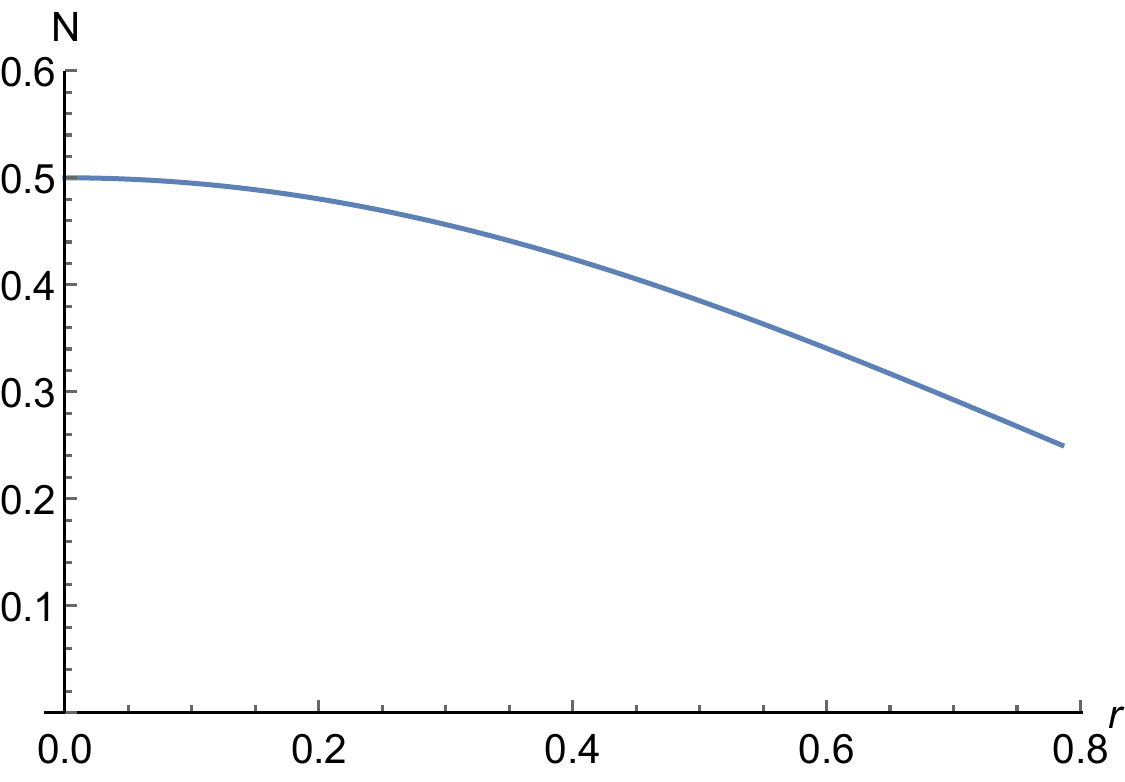}
\caption[]{\it  Negativity as a function of r.} \label{fig:3}
\end{figure}

There is a necessary condition stating that if there exists at least one negative eigenvalue of the partial transpose of a bipartite density matrix, the density matrix is entangled \cite{peres1996separability}. So, negative eigenvalues of the partial transpose of a density matrix can be a good indicator of how a system is entangled. Based on this criterion, negativity, may be introduced as in Eq.(\ref{negativ}) below as an entanglement measure distinguishing between the different bi-partitions of a system \cite{vidal2002computable}
\begin{equation}\label{negativ}
N_{AB}=\sum_{i} \frac{\mid \lambda_{i}\mid -\lambda_{i}}{2}=-\sum_{\lambda_{i}<0} \lambda_{i} ,
\end{equation}
where, $\lambda_{i}$s are the eigenvalues of a partial transpose of the bi-partite density
matrix $\rho_{AB}$. In order to find the partial transpose of the density matrix of our system $\rho_{A,I}$ (Eq.(\ref{eq:11})), we need to interchange Alice's qubits $(\mid a_{A},b_{I}\rangle\langle c_{A},d_{I}\mid\longrightarrow \mid c_{A},b_{I}\rangle\langle a_{A},d_{I}\mid)$. The negative eigenvalues of the partial transpose of $\rho_{A,I}$  are $ \lambda_{i}= \lbrace \frac{-\beta^{2}(1+cos(2r)}{4},\frac{-\alpha^{2}(1+cos(2r))}{4}\rbrace$.
Therefore, the negativity may be evaluated along the following lines:
\begin{eqnarray}
N(\rho_{A,I}) &=& \frac{1+ (\alpha ^{2}+ \beta ^{2})\cos(2r)}{4}. \nonumber\\
 &=&\frac{\cos^{2}(r)}{2}.
\end{eqnarray}
Negativity does not depend on $\alpha$ and is a decreaseing function of $r$.
The negativity is plotted for different values of $r$ in Fig.(\ref{fig:3}).
\\
\section{\label{sec:level4}Conclusion}
Considering a maximally entangled Bell sate with two modes detected by Alice and Rob, we proposed a new form of vacuum state for Rob which has an entanglement between the momentum modes leading to an excited state as in Eq.\eqref{eq9c} and we studied the entanglement of this system. We realize that entanglement of this system does not depend on the amount of entanglement of the vacuum state but it decreases with increasing the acceleration.

In this paper, we just studied the entanglement between the modes $A$ and $I$. Other cases like entanglement between the modes $A$ and $II$ while those of $I$ and $II$ could also be considered.
It is also interesting to quantify other correlation quantifiers such as quantum discord and mutual information.
\section{Acknowledgements}
We would like to thank Mrs Zeynab Harsij for her comments.
\appendix
\section{Purity}
\begin{figure}[h]
\centering
\includegraphics[width=0.4\textwidth]{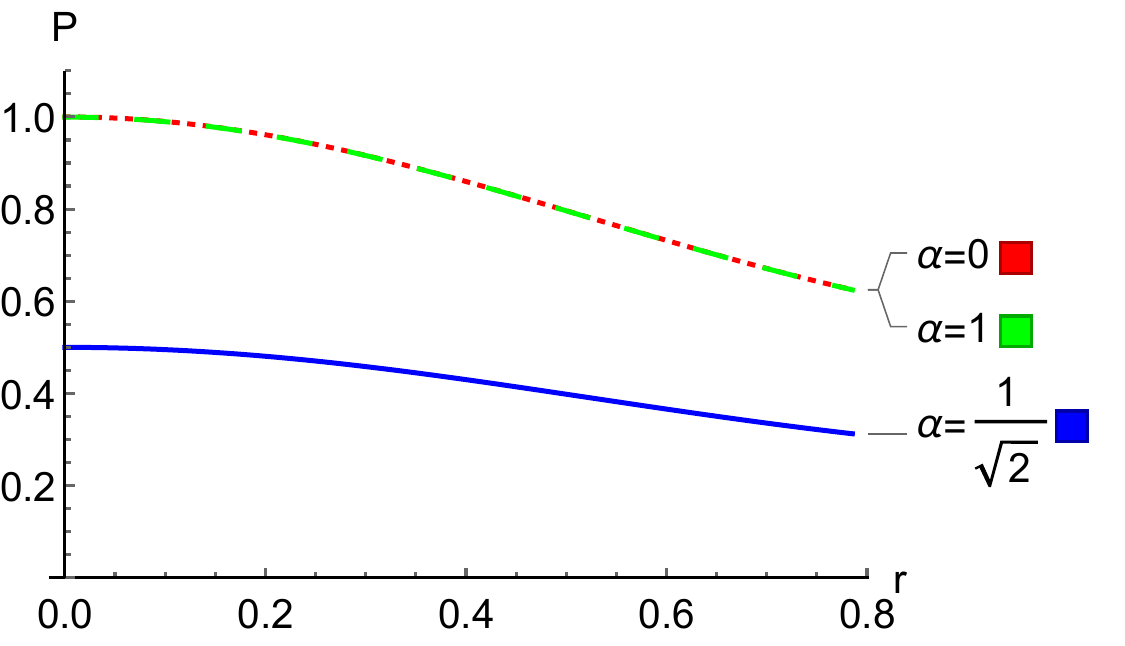}
\caption[]{\it Purity as a function of $r$ for some values of $\alpha$.\\Dashed curve (Green) for $\alpha=1,\beta=0$. Thick solid curve (Blue) for $\alpha=\beta=\frac{1}{\sqrt{2}}$. Dotted curve (Red) for $\alpha=0$,  $\beta=1$.} \label{fig:4}
\end{figure}

It was shown that the entanglement between the $k$ and $-k$ modes does not affect the negativity. We explored the potential effects of this proposal on other correlations of the system. For this purpose, we computed the two other measures of purity and relative entropy of coherence. Purity is a measure representing the degree of mixedness of states \cite{nielsen2000quantum}. The concept is defined as follows:
\begin{equation}
P=Tr[\rho^{2}].
\end{equation}
As expected, purity should be equal to unity for pure states and minimum for maximally mixed states (Fig.(\ref{fig:4})). Compared to other values of $\alpha$ in the case of $\alpha=\frac{1}{\sqrt{2}}$, we have a more stable value for purity when acceleration changes.
\section{Relative entropy of coherence}
\begin{figure}[h]
\centering
\includegraphics[width=0.4\textwidth]{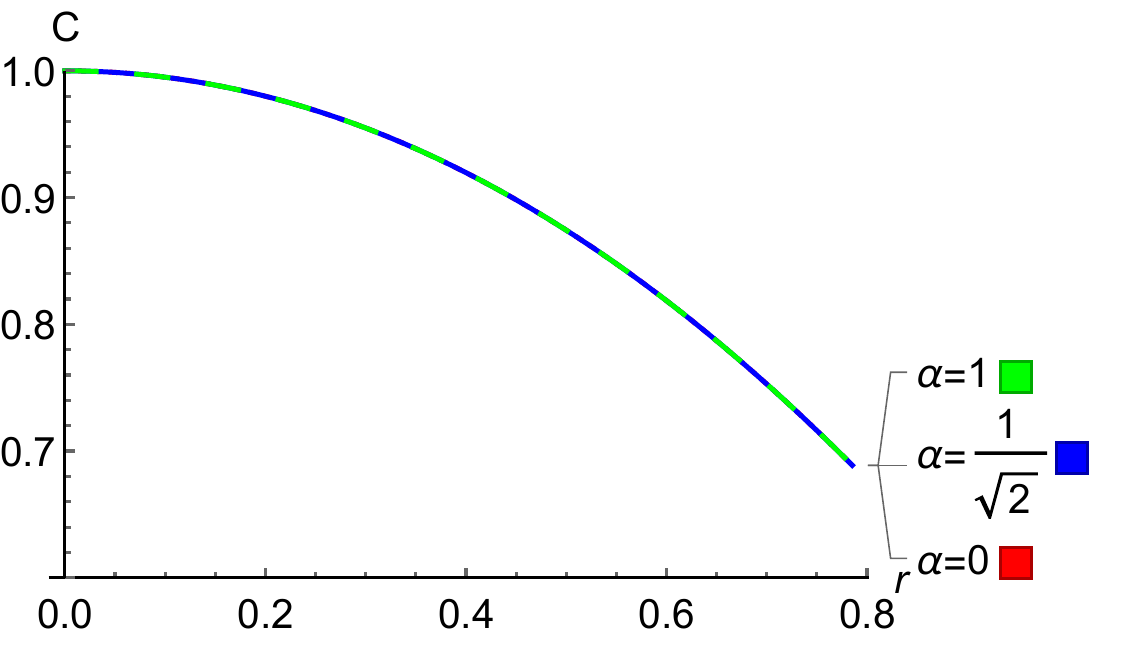}
\caption[]{\it ( Color online) Relative entropy of coherence as a function of r. Thin dashed curve (Green) for $\alpha=1,\beta=0$. Thick solid curve (Blue) for $\alpha=\beta=\frac{1}{\sqrt{2}}$. Dotted curve (Red) for $\alpha=0,\beta=1$. } \label{fig:5}
\end{figure}

 Coherence is an important subject in optics and quantum mechanics and plays a central role in interference. In fact, coherence describes all the correlation features between two waves. In quantum mechanics, we can find the coherency of a state based on its density matrix. One such quantifier is relative entropy of coherence
\cite{baumgratz2014quantifying}:
\begin{equation}\label {eq:19}
C_{r}(\rho) := S(\rho_{diag})-S(\rho) ,
\end{equation}
where, $\rho_{diag}=\sum\rho_{ii}\mid i\rangle\langle i\mid$ which is obtained by removing off-diagonal elements of $\rho$. The state is incoherent when $C_{r}(\rho)= 0$. Using Eqs. (\ref {eq:11}) and (\ref {eq:19}), the relative entropy of coherence is obtained and plotted in Fig. (\ref{fig:5}).
It is seen in the Fig.(\ref{fig:5}) that the relative entropy of coherence is nonzero, which means that this system is always coherent except for the case of $\alpha=0$. When acceleration is zero, coherency vanishes and the system becomes incoherent. As we can see coherency has the same value for all values of $\alpha$ and it decreases with increasing the acceleration.

{}

\end{document}